# Synthesis of Single-Crystalline Lead Sulfide Nanoframes and Nanorings

*Sascha Kull,[1] Leonard Heymann,[1] Ana B. Hungria,[2] Christian Klinke[1,3,4]\**

[1] *Institute of Physical Chemistry, University of Hamburg,*
*Martin-Luther-King-Platz 6, 20146 Hamburg, Germany*
[2] *Universidad de Cádiz. Facultad de Ciencias,*
*Campus Rio San Pedro, Cadiz 11510, Spain*
[3] *Department of Chemistry, Swansea University – Singleton Park,*
*Swansea SA2 8PP, United Kingdom*
[4] *Institute of Physics, University of Rostock,*
*Albert-Einstein-Straße 23–24, 18059 Rostock, Germany*

**Abstract**
We present a colloidal synthesis strategy to obtain single-crystalline PbS nanorings. By controlling the ripening process in the presence of halide ions, a transformation of initial PbS nanosheets to frame-like structures and finally to nanorings was achieved. We found that the competing ligands oleic acid, oleate and halide ions play an important role in the formation of these nanostructures. Therefore, we propose a formation mechanism based on a thermally induced ripening of crystal facets dependent on the surface passivation. With this method, it became possible to synthesize colloidal nanorings of cubic crystal phase galena PbS. The synthesis was followed via TEM and the products are characterized by XRD, AFM and STEM tomography. Control of the initial nanoframe morphology allows adjusting the later nanoring dimensions.

\* Corresponding author: christian.klinke@uni-rostock.de

**Introduction**

In the last decades, a large variety of colloidal nanomaterials were developed. Beginning with zero-dimensional (0D) nanoparticles of metals and semiconductors of different shapes (e.g. spheres, cubes, octahedrons, stars), it is now also possible to synthesize one-dimensional (1D) nanorods and nanowires of different morphology and two-dimensional (2D) nanoplatelets and nanosheets.[1,2] Among these, nanorings are of interest in terms of quantum phenomena to appear when electron or hole wave functions are confined in a toroidal system. A magnetic field induced transition from ground state to chiral state was shown for MBE (molecular beam epitaxy) grown InGaAs nanorings[3] and theoretical works suggest interesting features of nanorings like Aharanov-Bohm effect.[4–7] Metallic nanorings of silver,[8] gold,[9–11] nickel[12] and cobalt[13] were produced by usage of lithography techniques involving a template, metal deposition and argon ion etching. These nanorings were studied in terms of electrical and optical properties[11] and possible applications in catalysis,[14] sensors[8,9] and quantum-mechanical devices.[13] The fabrication of the studied nanoring devices often requires complex, expensive and time-consuming instrumentation. In contrast, colloidally synthesized nanostructures have the advantage of using simple instrumental set-up, low-cost chemicals, scalable syntheses, and effortless processing. However, colloidal nanorings cannot be obtained easily as the morphology is thermodynamically not favorable, but strategies have been developed and applied to achieve this complex morphology in colloidal nanocrystal syntheses. All reports in literature follow the approach of modifying a starting nanostructure (e.g. by etching) to achieve a hollow or ring-like morphology as shown for gold nanostructures,[15] Pd and Pt nanorings[16,17] and PbS nanospheres.[18] Another attempt is a controlled nanoparticle aggregation to hollow spherical superparticles[19–21] or nanorings.[22] A hollow sphere or ring-like morphology can be achieved but the nanostructures have a low crystallinity and possess grain-boundaries. More promising approaches are based on single-crystalline 2D nanomaterials[23], by usage of the Kirkendall-effect on CuSe/CuS core shell nanoplatelets,[24] via oriented attachment[25] nanosheet templated etching[26,27] or rearrangement[28] to nanorings.

We devote our study to lead(II) sulfide, a prominent narrow-band gap semiconductor with an $E_g$ of 0.41 eV.[29] It crystallizes in the cubic rock salt structure with the space group *Fm-3m*. The exciton-Bohr radius is 20 nm with similar masses for electrons and holes, enabling strong quantum confinement effects when at least one dimension is minimized.[30] In general, the colloidal synthesis of 1D or 2D PbS is challenging and cannot be achieved by simple ligand adjustments as long as the crystal is isotropic. Oriented attachment of initial seeds/clusters can be exploited to achieve these kinetically formed nanostructures when a soft template is present.[31] In the case of 2D PbS, oleic acid, which is known to form lamellar phases, acts as the soft template directing the attachment of initial particles into a two dimensional "egg tray-like structure". The arranged building blocks merge to a stable single crystalline nanosheet.[32] With the same approach PbS nanostripes were accessible based on a manipulation of the precursors reactivity by increasing the amount of the co-solvent from the halogenated hydrocarbons.[33]

Lead(II) sulfide nanomaterials are studied intensively in the last decades starting with nanoparticles[34,35] and their application in photovoltaics.[36–38] Further, PbS nanosheets were investigated as single-sheet transistor devices.[39,40] Multi-exciton generation and tunable electrical features by halide ion doping were studied to improve nanosheet based devices as possible components in semiconductor industries.[41–43]

Here, we present the synthesis of PbS nanoframes as intermediate nanostructures on the pathway to PbS nanorings. The synthesis starts with the formation of initial PbS nanosheets, which undergo a thermally induced ripening process in the presence of halide ions first forming nanoframes and later nanorings via additional halide ion etching. The nanostructures are characterized by transmission electron microscopy (TEM), X-ray diffraction (XRD) and atomic force microscopy (AFM). A formation mechanism is proposed explaining the formation of the complex PbS nanostructures from the highly symmetric cubic galena phase. Control of the reaction conditions allows tuning the nanoframes morphology in terms of lateral dimensions and thickness and with this, the later nanorings morphology.

**Experimental section**
*PbS nanoring synthesis*
All chemicals were used as received. The following chemicals were used: 1-bromtetradecane (BTD; Aldrich, 97%), diphenylether (Sigma-Aldrich, 99%), lead(II) acetate trihydrate (Sigma-Aldrich, 99.999%), lead(II) oxide, lithium bromide (Merck, <=100%), lithium chloride (Sigma-Aldrich), lithium iodide (Merck, >= 98.0 %), N,N-dimethylformamide (DMF; Sigma-Aldrich, 99.8% anhydrous), oleic acid (OA, Sigma-Aldrich, 90%), heptylamine (Sigma-Aldrich, 99%), thioacetamide (TAA, Sigma-Aldrich, >99.0%), toluene (VWR, >=99.5%).

Lead(II) sulfide nanorings were synthesized as followed. In a three-neck flask with thermocouple, condenser and septum 860 mg lead(II) acetate trihydrate (2.3 mmol), 3.5 mL oleic acid (10 mmol) and 10 mL diphenylether were mixed and degassed at 70 °C in vacuum for at least 2 h. Under nitrogen atmosphere, the solution was heated to 220 °C. During the heating step, 0.05 mL of a 0.15 M LiCl-DMF solution was added. When the reaction solution reached the desired temperature 0.2 mL of a 0.09 M TAA-DMF solution were added. The solution turned black immediately and showed metallic luster. At this moment PbS nanoframes were formed and have been etched to nanorings by addition of 100 µl BTD. The solution was stirred at 220 °C for additional 5 minutes after addition of the etchant and the black color turned brighter with time. When the reaction solution cooled down to room temperature, it was centrifuged at 4000 rpm (2186 rcf) for 3 min. The colorless supernatant was disposed and the black precipitate was washed two more times by dispersion in toluene and centrifugation as described above. The nanoframes and nanorings were stored in toluene and showed to be stable for several months.

*Characterization Methods*

AFM: Atomic force microscopy measurements were performed in tapping mode on a JPK Nano Wizard 3 AFM. The samples were prepared by drop-casting a diluted suspension on a silicon wafer.

TEM: The TEM samples were prepared by diluting the suspension with toluene followed by drop casting 10 µL of the suspension on a TEM copper grid coated with a carbon film. Standard imaging was performed on a JEOL-1011 with a thermal emitter operated at an acceleration voltage of 100 kV.

HAADF-STEM Tomography/ XEDS: Specimens for HAADF-STEM experiments were prepared by depositing particles of the samples to be investigated onto a copper grid supporting a perforated carbon film. Deposition was achieved by dipping the grid directly into the powder of the samples to avoid contact with any solvent. Both tomography and XEDS experiments were performed in a FEI Titan3 Themis 60–300 Double Aberration Corrected (AC) microscope operated at 200 kV. For the tomography studies, a convergence angle of 9 mrad was selected in order to improve the depth of focus and a camera length of 115 mm was used. The software FEI Explore3D v.4.1.2 facilitated the acquisition of the tomography tilt series from −64° to +76° every 2° and the alignment and reconstruction of the data set. Avizo software was used for visualization. The compositional maps were performed by X-ray Energy Dispersive Spectroscopy in STEM mode (STEM-XEDS) using the SuperX G2 detector. Very high spatial resolution STEM-XEDS maps were acquired using a high brightness, sub-angstrom diameter, electron probe in combination with a highly stable stage, which minimized sample drift. Element maps were acquired with a beam current of 80–120 pA and a dwell time of 157 ms, which results in a total acquisition time of approximately 13 min.

XRD: X-ray diffraction measurements were performed on a Philips X'Pert System with a Bragg-Brentano geometry and a copper anode with a X-ray wavelength of 0.154 nm. The samples were measured by drop-casting the suspended nanosheets on a <911> or <711> cut silicon substrate.

Binding energy simulations: In order to evaluate the adsorption energy of various halogen ions on the PbS facets, we performed simulations using DFT in the frame of the ORCA[44] software. We used the LDA exchange functional, the correlation functional VWN-5[45], the effective core potential LANL(78)[46] for Pb, and the def2-TZVP[47] basis set. The adsorption energy was calculated by comparison of the sum of the energies of the PbS crystal and the halogen ions as separated moieties, and the total energy of the combined, optimized system, in which the crystal geometry was kept fixed at the experimental values and the ions were free to relax.

**Results and Discussion**

The PbS nanoring synthesis can be described in three steps starting with the nanosheets formation. These steps are illustrated in Figure 1 with TEM pictures and a three-dimensional (3D) model of the PbS crystal, representing the different crystal surfaces. The oriented attachment of initial nanoparticles, leading to 2D nanostructures, is well described in literature.[32,48] The 2D PbS nanosheets ripen thermally induced in the presence of halide ions to nanoframes, which is the first step. PbS nanoframes have the same crystallographic orientation as PbS nanosheets[33] with <100> z-axis and edges in <110> direction (Figure S1). Additional halide-ion etching is performed to produce nanorings in the second step, while control of this etching step is necessary to prevent the third step, which leads to fragments and small particles.

The formation of frame-like nanostructures from nanosheets (Figure 1: step 1) is not trivial but can be explained by common phenomena in nanochemistry. It is well known, that the nanoscale crystals undergo a transformation to a three-dimensional (bulk) crystal in order to avoid a thermodynamically unfavorable surface-to-volume ratio. This is achieved when the nanostructures are allowed to grow further under thermodynamic conditions, e.g. when high temperatures are applied to a system over a longer period under growth conditions. When this thermodynamic growth is applied to kinetically formed products like PbS nanosheets, control of the process can achieve non-trivial nanostructures. The growth process can be controlled by temperature and reaction time or a ligand mediated change in crystal facet energies leading to a change in expressed crystal surfaces.

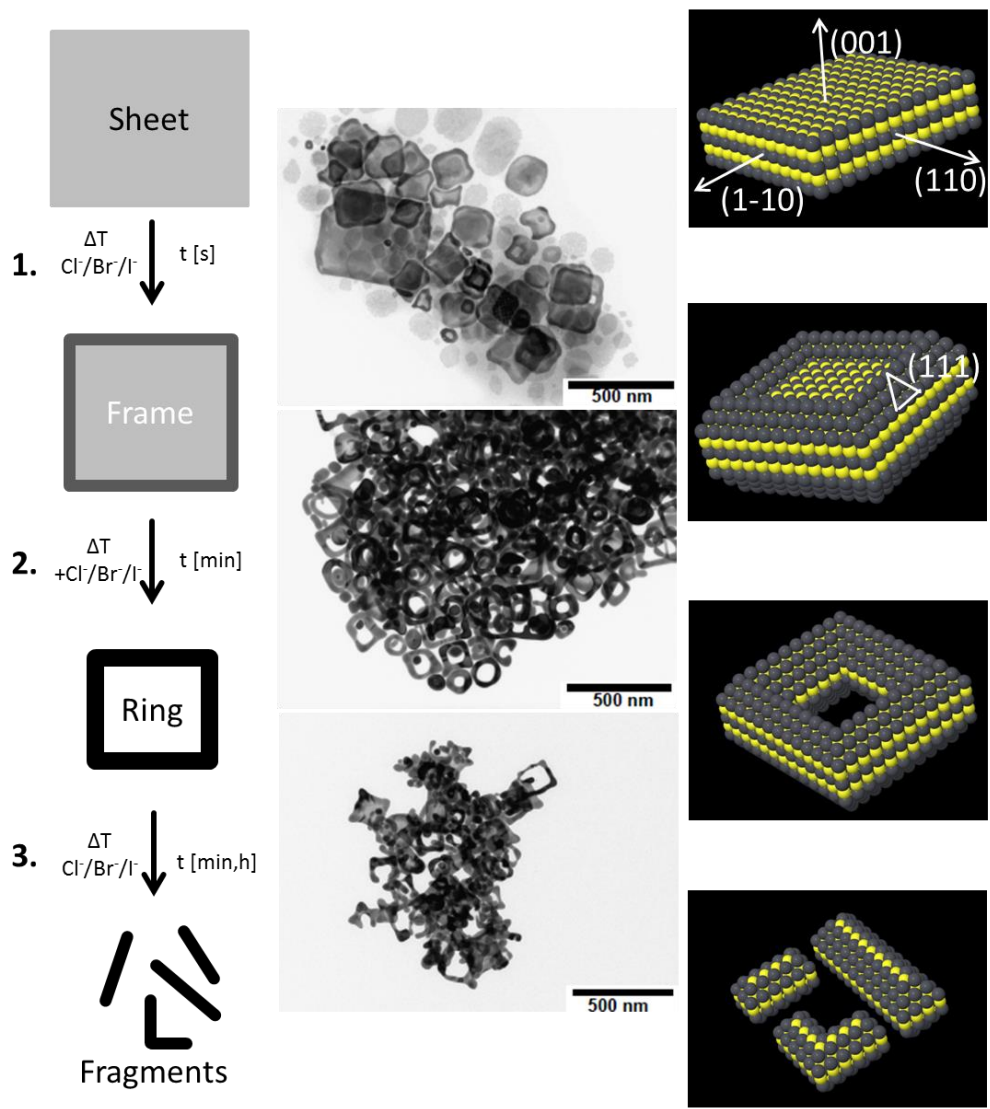

*Figure 1: PbS nanosheets evolve to nanoframes (1.) when halide ions are present at temperatures above 200 °C, while further halide ion etching leads to nanorings (2.), which decompose to fragments when the synthesis is prolonged further (3.). Crystallographic models of the nanostructures build up by lead (grey) and sulfur (yellow) atoms, indicating the different low index facets.*

Our approach bases on oleic acid as a soft template layer directing the PbS nuclei to attach two-dimensionally, avoiding growth in the third dimension. This passivation of a dense layer of oleic acid or lead oleate is affected by the temperature in a way that the adsorption/desorption of ligands is accelerated when temperature is raised. While individual ligand molecules are stabilized in the ligand layer by van-der-Waals interactions of surrounding molecules, an increase of temperature and with this the vibrational energy, destabilizes this ligand layer. An increase of the temperature favors desorption of the oleic acid ligand layer resulting in growth[49] to thicker and laterally smaller nanostructures (Figure S2).

Since we use a hundredfold excess of lead(II) oleate compared to the sulfur source, we do not expect a change in the concentration by lead(II) sulfide precipitation to play a role. Instead, the micellar phase and the templating effect of the lead(II) oleate in diphenylether (DPE) can be addressed by a change of excess of oleic acid or the solutions viscosity (e.g. by temperature). When the conditions are chosen in a range of optimal values, nanosheets can be grown thicker starting from their edges, while the inner part of the sheets remains stable by the oleic acid layer protection. In this set of synthesis parameters, nanoframes can be isolated as synthesis product or further be modified by halide ion etching resulting in nanorings. Determination of the lateral dimensions of the PbS nanostructures in TEM images revealed first hints regarding the unusual three dimensional topology of the nanoframes. They typically show edges with darker contrast in TEM due to an increase in thickness (Figure 2A). AFM measurements (Figure 2B) and high angle annular dark field scanning TEM (HAADF-STEM) tomography were used to investigate the complex height profile. HAADF-STEM tomography slices through the reconstructed volume (Figure 2C) and perpendicular (Figure 2D) were performed to verify the frame-like structure (additional data is provided as a video in the SI). HAADF X-ray energy dispersion spectroscopy (XEDS) was used to visualize the element mapping of lead (Pb) and sulfur (S) in the nanoframes and shows a homogenous distribution of both elements (Figure 2E).

To investigate the ripening process we performed syntheses with altered conditions. We obtained nanosheets instead of nanoframes when the amount of oleic acid (OA) is increased: e.g. 7 mL OA (2x default value) or 5 mL OA together with 0.2 mL sodium hydroxide to keep the oleic acid/oleate ratio constant. Both syntheses show no hints for a frame-like morphology in TEM or AFM (Figure S3), but a frame-like morphology can be obtained with these oleic acid amounts when the amount of chloride ions (as lithium chloride) is increased at the same time. This clearly shows the chloride ions to enable a local growth in height resulting in the frame-like morphology.

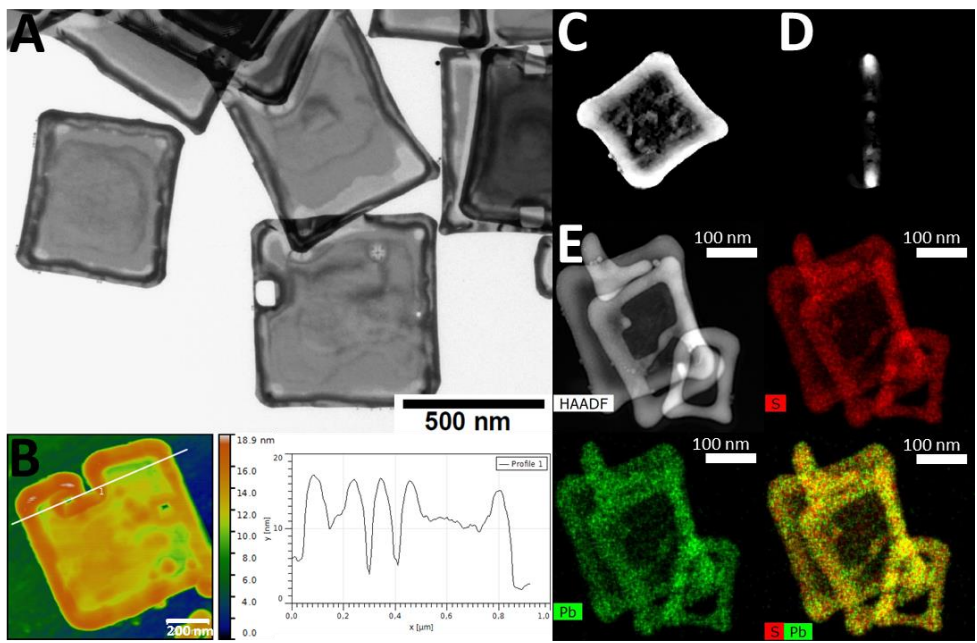

*Figure 2:* TEM image of synthesized nanoframes (A), with AFM image and corresponding height profile (B), representing the unusual height profile of nanoframes. STEM tomography slices through the reconstructed volume, parallel (C) and perpendicular to the nanoframes (D) and HAADF-STEM XEDS element mapping of an ensemble of nanoframes (E).

The role of chloride ions in the synthesis can be explained by the need for small ionic ligands for surface passivation. Long chained ligands result in an insufficient passivation because of the steric hindrance of the hydrophobic alkyl chain and therefore higher surface energy.[50] For PbSe, computer simulations were performed to predict nanoparticles faceting. It was found that oleic acid/oleate preferably binds to the {111} facets.[51] We calculated the binding energy of ionic ligands like oleate, acetate[48] or halide ions (table S1) to the {100}, {110} and {111} facets and found it to be the highest for the lead terminated {111} facet due to electrostatic interactions with increasing uncompensated charges at the surface.

When no halide ions are present at all, for example amines can be used as the second ligand[52], again normal nanosheets are formed (Figure 3A), which start to decompose when the temperature is raised above 200 °C. This leads to 2D based irregular shaped structures with holes, but without thicker grown edges (Figure 3B). The simple addition of lithium chloride directly after the nanostructure formation step, leads to the same 2D nanostructures but with thicker edges this time (Figure 3C). Again, the appearance of thicker edges of the 2D material can be directly linked to the presence of chloride ions after the sheets are formed. It should be noted that the system has to be in the growth stage at that time. A stable and not growing system would not result in frame formation, independent of the presence of halide ions, since there is no mass transfer to the edges.

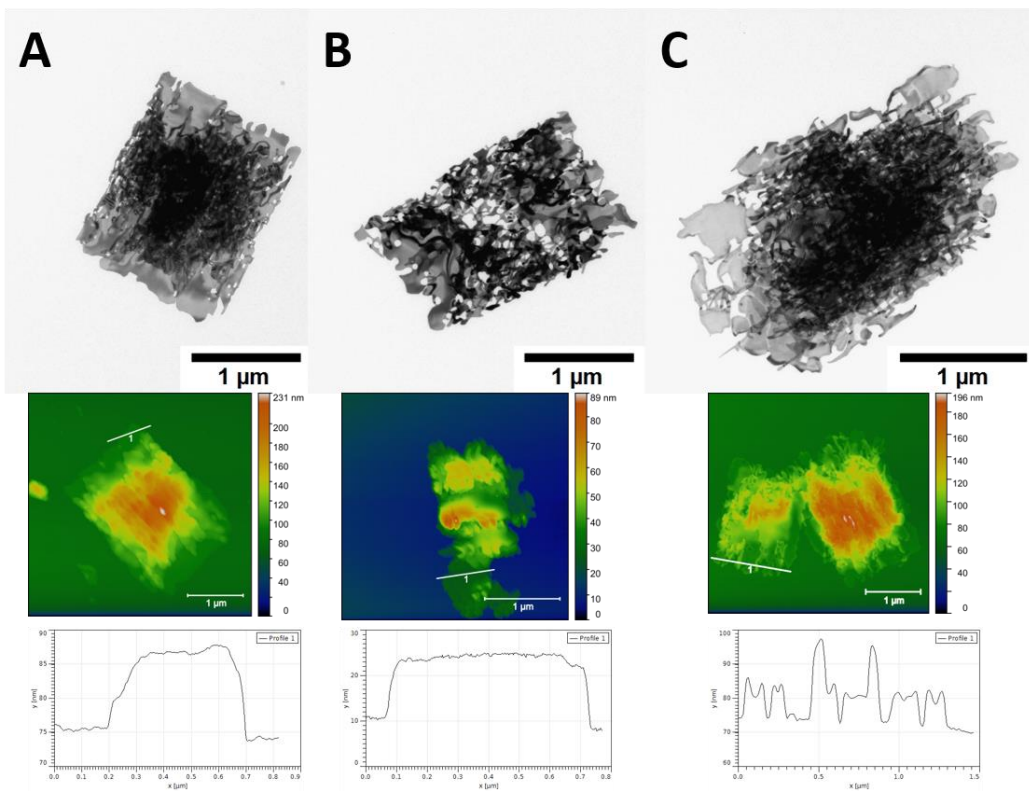

*Figure 3: Nanosheets obtained without any halogen compound but heptylamine (A), after temperature raise (B) and the same with lithium chloride addition after the nanosheets formation (C) leading to nanoframes this time. The ripening produces thicker edges of the initial nanosheets, as observable in the corresponding AFM height profiles.*

With oleic acid, lead(II) oleate and LiCl used as ligands in our system, the chloride ions are supposed to fulfil two tasks. First, to trigger the oriented attachment to 2D nanostructures and second, to compete with OA at the edges to promote growth in height. We expect lead terminated {111} facets to occur, which need additional passivation by smaller molecules for charge compensation.[53] Oleate and chloride ions terminate the lattice[54] resulting in an overall charge compensation when passivation is stoichiometric. Beside the passivation of the lead rich {111} facets, the halide ions accelerate the dissolution of PbS. This can be observed by a faster decomposition of nanosheets to fragments when the temperature is raised to a certain threshold temperature or the reaction time is extended to several hours. This halide ion promoted decomposition is used in the etching of nanoframes to nanorings (Figure 1, Step 2), which is described next.

For the etching of nanoframes to nanorings additional halide ions are necessary following the reactivity series Cl$^-$ < Br$^-$ < I$^-$. In the case of chloride ions, the etching to rings is observed at fewer nanoframes, while comparable amounts of bromide ions lead to a higher yield of nanorings. Iodide ions are more reactive and the ripening already leads to sticklike fragments when comparable conditions were used (Figure S4). The etching resulted in nanorings as well as wire fragments and nanoparticles, as can be seen in all TEM images and is illustrated in Figure 1. The synthesized PbS nanostructures were characterized by XRD. The galena phase was verified, and the lead halide phases were excluded in the synthesis product (Figure S4A).

We note that, when nanoframes are etched to nanorings, this etching always leads to a growth at the edges resulting in thicknesses in the range of 25-30 nm. The balance of dissolution and growth at the same time but at different locations, dependent on the surface energy, is the key to achieve ring-like nanostructures, not the etching in terms of a pure dissolution reaction. We additionally performed post-synthetic room temperature etching with and without halide ions and found this to be driven by the solubility of PbS in the reaction solution. The dissolution of PbS galena depends on the crystallographic facets and the rate is increasing from the {100} to the {110} to the {111} facets.[55] Adjustment in polarity and solubility of PbS by ligands shifts the equilibrium to the dissolution reaction and the nanoframes dissolve everywhere, meaning not selectively in the central region of the initial sheets. Therefore we refer to the here presented process as a thermal-induced ripening and we speak of "etching" only to differentiate the first and second step in our mechanism.

We found that the etching with halide ions to nanorings works best by addition of bromide ions in the thermodynamic growth stage. Halide ions penetrate the oleic acid/oleate layer and coordinate lead cations at the surface by a simple ligand exchange mechanism.[56,57] This results in a destabilization of the {100} top and down facets, which additionally are thinner than the already thicker grown edges. Further optimization of the etching process led us to 1-bromotetradecan (BTD) as etching solvent to produce the best samples of etched PbS nanorings so far. BTD releases bromide ions with time, which possess a moderate etching strength, and additionally act as long-chained ligand to penetrate and destabilize the oleic acid layer at the {100} top/down surface. When the etching is prolonged or harsh conditions are used, stick- and wire-like fragments are formed (Figure S4E). This finding suggests a change in chemical environment, which is inflicting the oleic acid soft template.[31] OA is known to show different aggregation morphologies, like micelles and lamellar phases, under variable chemical environment.[58,59] For example, Janke et al. found in their simulations that in water OA undergoes an evolution of lamellar to toroidal and worm-like shaped micelles, depending on concentration and pH value.[60] This behavior may serve as additional argument for the presence of a templating effect of OA but with curved morphology on the surface of the toroidal edge of PbS nanorings. Therefore, we assume the salt addition in our synthesis to influence the polarity of the medium and that of the micellar soft template, which has a strong impact on the ripening process.

In general, to synthesize nanoframes and later nanorings with a uniform morphology and thickness, the starting material, namely PbS nanosheets, has to fulfill crucial requirements with the priority as follows: (1) Homogenous surface; (2) Defined shape of the edges; (3) Uniformity in lateral dimension. These points are discussed further individually. In our presented synthesis we exchanged the halogenated hydrocarbon with lithium halide salts dissolved in *N,N*-dimethylformamide (DMF), which is already introduced to the synthesis as solvent of the sulfur source thioacetamide (TAA). We found halide ions to promote the oriented attachment leading to nearly the same nanostructures under reported conditions[52] and suggest the release of halide ions from the co-solvent to play an important role in the synthesis. However, this unpredictable side reaction, which is influenced by several parameters (e.g. temperature, time and concentration), can be excluded by the addition of defined amounts of lithium halides to the synthesis. We exchanged lead acetate to lead oxide to exclude residual acetate influencing the synthesis of nanoframes, like shown for PbS nanosheets.[61] In fact, we found a slight change in reactivity but the resulting nanoframes look the same for both lead precursors with no major differences. Therefore, we exclude remaining acetate or acetate formed during the decomposition of TAA to play a significant role in our mechanism.

Usually nanosheets synthesized at lower temperature (e.g. 130 °C) show strong stacking of the formed sheets and growth of additional layers starting from the interior.[48] When the temperature is raised higher than 180 °C, we found homogenous flat surfaces to become dominant. When the temperature is raised further to 220 °C and higher, laterally smaller nanosheets (respectively nanoframes) were formed resulting in a narrower overall size distribution. We focused our work on the temperature range between 200 °C and 240 °C and tackled the problem of defined edges by OA/chloride ligand adjustments. The thickness evolution is contrary to what was found for PbS nanosheets[48] but can be explained by a change of the lamellar template (thin sheets) to toroidal micelles[60], while the already high reactivity of the lead precursor is not affected in this temperature range.

Nanoframes, which fulfill our requirements for etching to rings, were found at reaction temperatures of 220 °C and oleic acid amounts of 2.5, 3.5 and 4.5 ml. With increasing OA volume, the resulting nanoframes become larger in lateral size, allowing tuning the resulting nanorings in the range from 300 nm to 1 µm (Figure 4). The absolute values of OA represent a change in excess oleic acid compared to the constant amount of OA used to form the lead(II) oleate precursor. With 2.5 ml OA the ratio of "free" OA to Pb bound oleate is 0.7, changing to 1.4 and 2.1 when 3.5 ml and 4.5 ml are used.

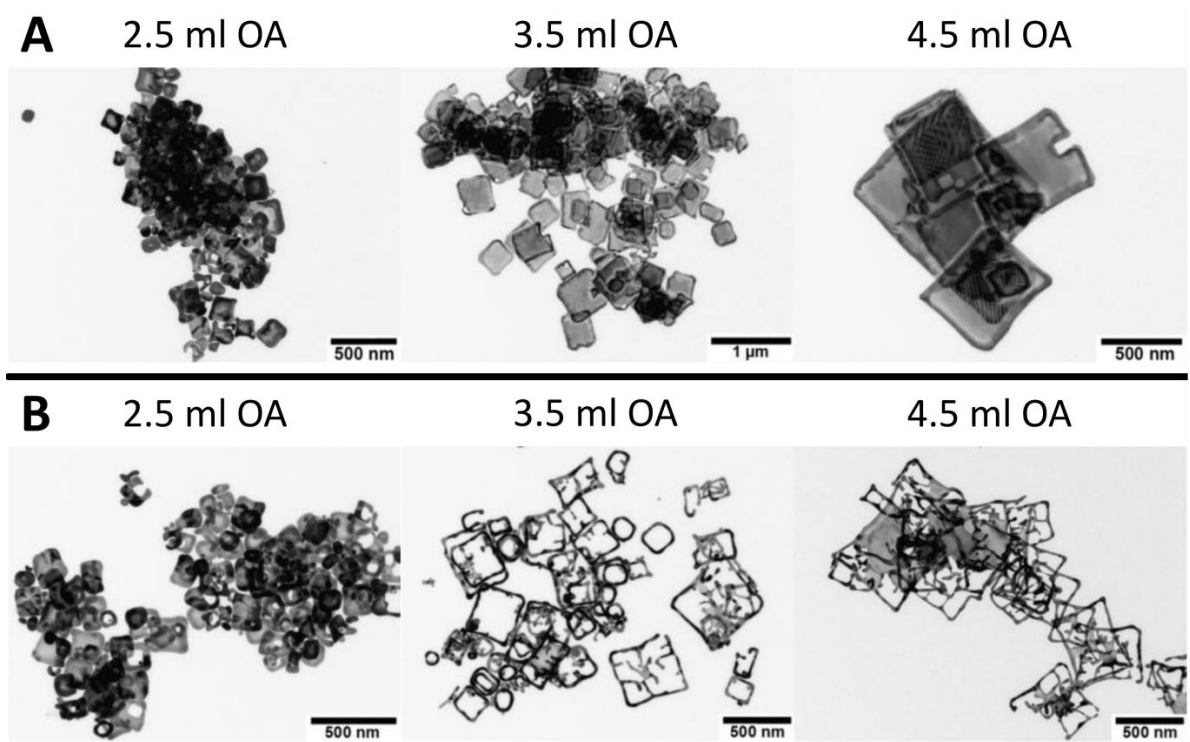

*Figure 4:* TEM images of nanoframes (A) synthesized with increasing amount of oleic and the resulting nanorings after etching for 5 min with 100 µl BTD (B).

The thickness of the nanoframes in the center decreases at the same time from 12.5 to 9.5 and 6.5 nm (determined by XRD) and is confirmed by AFM (Figure 5A). The change in thickness has a drastic influence on the etching of nanoframes to nanorings. When nanoframes grow thicker very quickly, as observed for 2.5 mL OA or higher temperatures (Figure S2), the etching with halide ions was not successful as the difference of height from the thicker edge to the interior of the frames is smaller. On the other hand, with 4.5 mL OA, the etching works quite well as the difference in height is more pronounced, though it yields samples that are more inhomogeneous. After etching, all nanostructures possess a height of 25 to 30 nm measured by AFM, confirming our idea of a thermally induced ripening with dissolution and growth taking place at the same time but different locations of the nanoframes (Figure 5B). Dissolution of the center part generates monomers, which then likely promote growth in height at the edges. Synthesizing nanoframes, which fulfill our requirements for later etching, was most promising at 220 °C with 3.5 mL oleic acid resulting in lateral dimensions of about 500 nm and resulted after a moderate etching step with BTD in the best sample of nanorings.

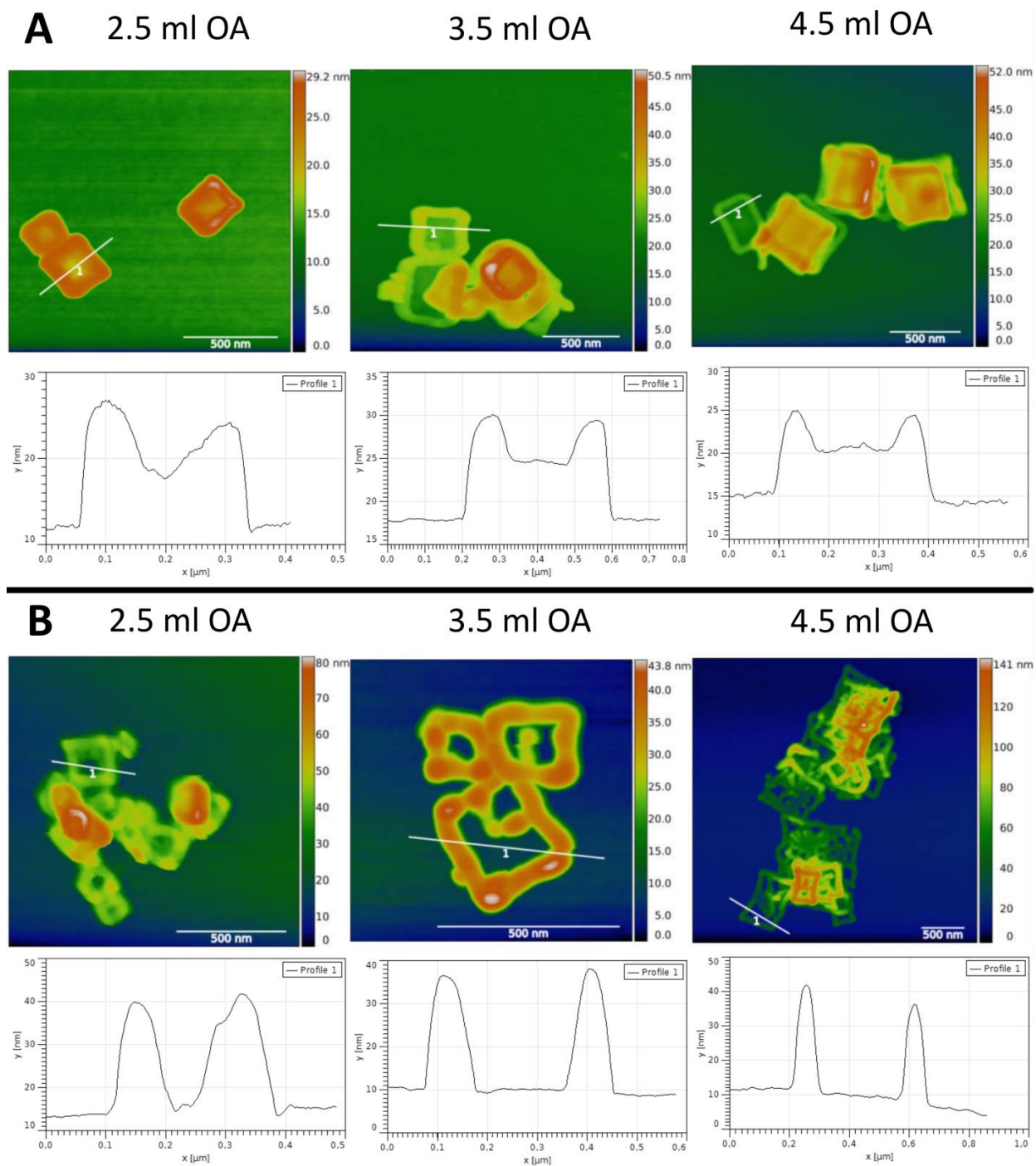

*Figure 5:* Corresponding AFM measurements of PbS nanoframes (A) with 2.5, 3.5 and 4.5 mL oleic acid in the synthesis, representative height profile underneath (location indicated by a white line in figure) and resulting nanorings (B) after etching with 100 µl BTD.

**Summary**

In summary, we developed a colloidal synthesis strategy to achieve single-crystalline toroidal nanostructures of cubic phase PbS. Based on the synthesis of PbS nanosheets we make use of a thermally induced ripening process to form nanoframes in a first step followed by an halide ion etching to nanorings. We demonstrate that the halide ions added in the synthesis play a crucial role in frame formation and the etching process on the road to PbS nanorings. The destabilization of the {100} facets is the key to control the thermal decomposition of PbS nanosheets to frame-like and even ring-like nanostructures. We found a change in surface passivation and therefore overall energy and stability of the facets when halide ions are added to the synthesis. This results in a growth in height at the edges, while the {100} facets are prone to thermal induced decomposition. Depending on the initial nanosheets and nanoframes, the morphology of the later nanorings can be adjusted in terms of the lateral size, while the etching always leads to thicknesses of >25 nm. The synthesized structures represent an interesting material for future devices based on quantum interferences in toroidal semiconductors.

**Acknowledgement**

C.K. thanks the German Research Foundation (DFG) for granting the project KL 1453/9-2 and the European Research Council is acknowledged for funding an ERC Starting Grant (Project: 2D-SYNETRA (304980), Seventh Framework Program FP7). A.B.H. thanks MINECO (Spain) for the project MAT2016-81118-P.

**Supporting Information Available:** Figure S1-S4, Table S1. A video based on HAADF-STEM tomography shows the three-dimensional structure of the nanoframes.